\documentclass[fleqn, usenatbib, useAMS]{mnras}
\usepackage{newtxtext,newtxmath}

\usepackage{graphicx}
\usepackage{amsmath}	
\usepackage{aas_macros}
\usepackage{natbib, twoopt}
\usepackage[utf8]{inputenc}
\usepackage{hyperref}
\usepackage{float}
\usepackage{multirow}
\usepackage{rotating}
\usepackage{tikz}
\usepackage{caption}
\usepackage{subcaption}

\hypersetup{colorlinks=true,linkcolor=blue,citecolor=blue,urlcolor=blue}

\bibpunct{(}{)}{;}{a}{}{,}

\makeatletter
\newcommandtwoopt{\citeads}[3][][]{\href{http://adsabs.harvard.edu/abs/#3}%
{\def\hyper@linkstart##1##2{}%
\let\hyper@linkend\@empty\citealp[#1][#2]{#3}}}
\newcommandtwoopt{\citepads}[3][][]{\href{http://adsabs.harvard.edu/abs/#3}%
{\def\hyper@linkstart##1##2{}%
\let\hyper@linkend\@empty\citep[#1][#2]{#3}}}
\newcommandtwoopt{\citetads}[3][][]{\href{http://adsabs.harvard.edu/abs/#3}%
{\def\hyper@linkstart##1##2{}%
\let\hyper@linkend\@empty\citet[#1][#2]{#3}}}
\newcommandtwoopt{\citeyearads}[3][][]%
{\href{http://adsabs.harvard.edu/abs/#3}
{\def\hyper@linkstart##1##2{}%
\let\hyper@linkend\@empty\citeyear[#1][#2]{#3}}}
\makeatother

\title[Dust nucleation around Wolf-Rayet stars]{Smoke on the wind: dust nucleation in archetype colliding wind pinwheel WR~104}

\author[A. Soulain et al.]{
A. Soulain,$^{1,~4}$\thanks{E-mail: anthony.soulain@univ-grenoble-alpes.fr}
A. Lamberts,$^{2,~3}$
F. Millour, $^{2}$
P. Tuthill, $^{4}$ and
R. M.\ Lau$^{5}$
\\
$^{1}$Univ. Grenoble Alpes, CNRS, IPAG, 38100 Grenoble, France\\
$^{2}$Universit\'e C\^ote d'Azur, Observatoire de la C\^ote d'Azur, CNRS, Lagrange, France\\
$^{3}$Universit\'e C\^ote d'Azur, Observatoire de la C\^ote d'Azur, CNRS, ARTEMIS, France\\
$^{4}$School of Physics, University of Sydney, NSW 2006, Australia\\
$^{5}$NSF’s NOIR Lab 950 N. Cherry Avenue, Tucson, AZ 85721, USA\\
}

\date{Accepted XXX. Received YYY; in original form ZZZ}

\pubyear{2022}

\begin{document}
\label{firstpage}
\pagerange{\pageref{firstpage}--\pageref{lastpage}}
\maketitle

\begin{abstract}
A handful of binary Wolf-Rayet stars are known to harbour spectacular spiral structures spanning a few hundred AU. These systems host some of the highest dust production rates in the Universe and are therefore interesting candidates to address the origin of the enigmatic dust excess observed across galactic evolution. The substantial interaction between the winds of the Wolf-Rayet star and its companion constitutes a unique laboratory to study the mechanisms of dust nucleation in a hostile environment. Using the grid-based \texttt{RAMSES} code, we investigate this problem by performing a 3D hydrodynamic simulation of the inner region of the prototypical spiral nebula around WR104. We then process the \texttt{RAMSES} results using the radiative transfer code \texttt{RADMC3d} to generate a candidate observable scene. This allows us to estimate the geometrical parameters of the shocked region. We link those quantities to the specific chemical pathway for dust nucleation, where the hydrogen-rich companion's wind catalyses dust formation. The scaling laws we derive constitute a unique tool that can be directly compared to observations. Depending on the dust nucleation locus, the velocity field reveals a differential wind speed. Thus, the initial dust speed could be more balanced between the speeds of the two stellar winds ($\sim$1600 km/s). With \texttt{RADMC3d}, we provide constraints on the dust nucleation radius for different combinations of dust-to-gas ratio, hydrogen enrichment and dust grain properties. Finally, our models reveal that dust may escape beyond the boundaries of the spiral due to hydrodynamical instabilities in the wind collision zone.
\end{abstract}




\section{Introduction}
Classical Wolf-Rayet (WR) stars are the immediate precursor stage to the supernova explosion of massive stars. Massive stars are thought to be responsible for much of the chemical evolution of galaxies, playing  their part in chemical enrichment and injecting kinetic energy that acts to moderate the stellar formation. WR stars play a crucial role in governing the proportion of observed carbon ($^{14}$C) or aluminum ($^{26}$Al) elements in the local interstellar medium, which are key ingredients of solar-like planetary systems \citepads{2010ApJ...714L..26T}. These direct descendants of massive O-type stars are characterised by a dense ($> 10^{-5}\,\dot{M}_{\odot}\,yr^{-1}$), fast ($> 1000$ km/s) and optically thick radiative wind that can entirely shroud the photosphere. Notwithstanding the violent conditions around WR stars that at first sight appear sterile for dust, a significant infrared excess emerges from spectroscopic determinations \citepads{1972A&A....20..333A, 1987A&A...182...91W}. Depleted of their hydrogen-rich shells, the early nitrogen-rich WR (WN subgroup) evolve into the late carbon-rich stage (WC subgroup) and, for some, initiate an era of efficient dust production. Despite their relatively short lifespan ($<$ 1 Myr, \citeads{2014A&A...564A..30G}) and them being one of the rarest type of stars in the Milky-Way ($\sim$670, \citeads{2015MNRAS.447.2322R}), WCs are thought to be a significant contributor to the galactic dust reservoir. In fact, decades of infrared surveys highlights an excess of galactic dust both locally \citepads{2009MNRAS.396..918M, 2012ApJ...748...40B, 2016MNRAS.457.2814S} and at high redshift \citepads{2014MNRAS.441.1040R}. Recently, \cite{2020ApJ...898...74L} revisited the impact of dust production from the Galactic WC population using a sample of 19 carbon-rich WR stars. Their SEDs comparison revealed a high dust production rate (DPR, $\dot{M}_d$$\sim$$10^{-10}$--$10^{-6}\,M_{\odot}/yr$) with a wide variety of carbon condensation fractions (0.002--40$\%$). Such DPRs can be compared with Galactic evolution models and thus could imply WC stars dominate galactic dust budgets at stages prior to the arrival of numerous low-mass evolved giant stars (AGBs). The case for extragalactic WC stars dominating dust production is even more compelling particularly in low-metallicity environments \citepads{2021ApJ...909..113L}. 

The Wolf-Rayet system WR~104 (MR80) represents the prototype of such dust producer. Initially, the presumption of dust formation was evoked to explain the large infrared excess observed by \citetads{1972A&A....20..333A}. The first attribution as a Pinwheel nebulae came from the first reconstructed image of the system by \citetads{1999Natur.398..487T} using Non-Redundant Masking (NRM) technique \citep{1986Natur.320..595B, 2000PASP..112..555T}. WR104 benefitted from intensive follow-up observations to understand its properties and constrain the orbital period (241.5 days) of its central binary \citepads{2008ApJ...675..698T}. In the early 2000's, space-based observations with the Hubble Space Telescope (HST) revealed a second companion ("companion B") at larger separation \citep{2002ASPC..260..407W}. A more recent study lead by our team confirmed the gravitational link with companion B but it appears to have no impact on the spiral shape itself \citep{2018A&A...618A.108S}. The distance to the system ($2.5\pm0.1$ kpc) is estimated following the method adopted by \citetads{2008ApJ...675..698T}. As the dust and gas are expected to be coextensive, the dust expansion speed is attributed to the terminal wind speed of the dominant WR star (1220 km/s measured by spectroscopy, \citealt{1992A&A...261..503H}). This assumption could be more complex than expected as recently highlighted by the important discrepancy detected between the spectroscopic and the dust velocities in the Apep system \citep{2020MNRAS.498.5604H}. An extensive study concerning the velocity field is therefore required to unravel the relationship between the gas and the dust grains involved in such binary systems.

Despite the hostile thermal and radiative environment, one mechanism for dust formation in WC stars is mediated by colliding-wind binary interactions, as proposed by \citetads{1991MNRAS.252...49U}. The apparent necessity to include a massive companion is consistently reinforced by both observations \citepads{2008ApJ...675..698T, 2009A&A...506L..49M, 2020MNRAS.498.5604H} and simulations \citepads{2012A&A...546A..60L, 2016MNRAS.460.3975H}. Importantly, massive stars are commonly (potentially always) involved in multiple systems \citepads{2013A&A...550A.107S}. A recent spectroscopic survey by \citeads{2020A&A...641A..26D} shows that 70$\%$ of the Galactic WR stars are in multiple systems. This suggests that a large fraction of massive stars could potentially present a dust production episode during their lifetime.

The wind collision zone (WCZ) generated by massive binary stars can protect the dust nucleation from the intense stellar UV radiation. Numerical simulations show that shocked winds are separated from each other by a contact discontinuity, which may be subject to various instabilities \citep{1992ApJ...386..265S}. The innermost hydrodynamic structure is imposed by the winds' momentum flux ratio and radiative cooling. In WR104, we anticipate the shock surface to be heavily skewed toward the O star companion, which boasts the much less dense (hence lower momentum) wind. Further out, orbital motion wraps the shocked structure into a spiral shape \citep{2012A&A...546A..60L}. While analytic estimates provide the location of the shocked region between the stars and the asymptotic opening angle of the WCZ \citep{1990FlDy...25..629L}, hydrodynamic simulations are essential to capture the complete 3D geometry of the interaction region. Only numerical simulations provide a comprehensive model of the density, velocity, and temperature conditions in the WCZ \citep{2009MNRAS.396.1743P}. Such efforts have been extensively used to constrain X-Ray observations \citep{2005xrrc.procE2.01P} or the gaseous structure of the dustless system $\gamma$~Vel by interferometry \citep{2017MNRAS.468.2655L}. \citet{2016MNRAS.460.3975H} investigates the effects of radiative cooling at larger scales, exhibiting a dramatic impact on the opening angle compared to previous studies. By including dust directly in their hydrodynamical modelling, they were able to retrieve the realistic density structure of the massive binary WR98a, a system similar to WR104, harbouring a pinwheel nebula \cite{1999ApJ...525L..97M}. Such spiral patterns have also been observed around massive red giant binary systems as recently modelled by \citep{2021MNRAS.507.4044C}. The binary interaction is also evoked to explain the 3D gaseous structures observed by ALMA around massive stars \citep{2012Natur.490..232M}.

Following these simulation efforts, we present our new \texttt{RAMSES} simulation of the very inner region of the system WR~104. We focus our attention on the first 50 AU where the dust is expected to form, as constrained by observation \citepads{2018A&A...618A.108S}. Our aims are to 1) understand the role of the companion in terms of hydrogen enrichment, 2) link the spiral morphology and the physics of the dust nucleation, 3) estimate the velocity map at the expected dust formation locus and 4) constrain the nucleation radius using radiative transfer modelling. The exquisite spatial resolution afforded by restricting the region of interest allows us to delve into the underlying physics of the dust nucleation processes. 

Our paper is organised as follows. In Sect. \ref{sec: hydro}, we discuss the numerical method and the initial and boundary conditions of the simulations. We present also the adopted methodology to compute the dust density grid. Section \ref{sec:setupRT} describes our radiative transfer approach applied to post-process the \texttt{RAMSES} simulation using the \texttt{RADMC3d} code. In Sect. \ref{sec:results}, we present the results focusing on the spiral morphology, velocity determinations and nucleation radius. In Sect. \ref{sec:discussion}, we interpret and discuss the astrophysical consequences of our results and finally, in Sect. \ref{sec:conclusion}, we conclude and place this effort into a wider astrophysical context.

\section{Hydrodynamical simulation}
\label{sec: hydro}
\subsection{Numerical setup and initialisation of the stellar winds}

We use the \texttt{RAMSES} code, which is based on a second-order Godunov method  \citep{2002A&A...385..337T} to numerically solve the system of hydrodynamics equations:
\begin{eqnarray}\label{eq:hydro}
    \frac{\partial\rho}{\partial t}+\nabla \cdot(\rho \mathbf{v})&=&0\\  \nonumber
	\frac{\partial(\rho \mathbf{v})}{\partial t}+\nabla (\rho \mathbf{v}\mathbf{v})+\nabla P 	&=& 0	\\
	  \frac{\partial E }{\partial t}+\nabla \cdot[\mathbf{v}(E+P)]	&=& n^2 \Lambda(T),	\nonumber
\end{eqnarray}
where $\rho$ is the density, $\mathbf{v}$ the velocity, $P$ the pressure of the gas and $n$ the number density. $\Lambda$ is the radiative cooling rate of the gas, based on \citet{1993ApJS...88..253S}. The cooling curve assumes an optically thin gas in ionisation equilibrium. We assume solar abundance for the O star wind and a WR abundance for the WR wind \citep{1992ApJ...386..265S}, which leads to increased cooling. We set the adiabatic index $\gamma=5/3$ and the total energy density $E$ is given by
\begin{equation}
E= \frac{1}{2}\rho v^2+\frac{P}{(\gamma -1)},
\end{equation}
To distinguish between both winds and quantify the mixing between them, we include a passive scalar, advected by each wind. To limit numerical quenching of hydrodynamic instabilities we use the MinMod slope limiter combined with the HLLC Riemann solver.

The simulation is based on a Cartesian grid with outflow boundary conditions. To study the dust condensation region, the simulation domain is 24 times the semi-major axis of the binary. The orbital plane is set at $z=0$ and we initialize the binary in the upper left corner of the domain. This allows us to model the interaction region with optimised computational cost. The simulation is stopped after 30$\%$ of one orbital period. The default resolution of the simulation is $N_x=256$ and we allow up to 5 levels of Adaptive Mesh Refinement (AMR) to locally increase the resolution for a limited computational cost. AMR is limited to 10$\%$ of the simulation domain above and below the mid-plane. The 3 highest levels of refinement (equivalent resolution of $N_x=2048$ and above) are restricted to smaller and smaller regions around the stars. The high resolution is necessary to insure spherical symmetry of the winds and guarantee that the shock region can develop naturally. Because of the strongly unequal winds, the shocks develop close to the O star and enough computational cells are necessary in this narrow region. Our physical resolution is about 0.007 AU, which should yield limited numerical heat conduction at the interface between the winds \citep{2010MNRAS.406.2373P}. 

The winds are generated following the same method as \citet{2011MNRAS.418.2618L, 2017MNRAS.468.2655L}. A density, pressure and velocity profile is set at each time step within a small region representing the star. Table~\ref{tab:simuhydro} shows the wind parameters for our simulation. The temperature in the winds are set to $10^5$K, which is much higher than expected in stellar winds. However, given the speed and density in the winds, this value guarantees that the winds are highly supersonic, in which case the shocked region is unaffected by the value of the pre-shock temperature. The orbital motion of the stars is determined using a leapfrog method. As the orbital velocity is negligible with respect to the speed of the winds, each wind can be considered isotropic in the frame co-rotating with the corresponding star.   

\begin{table}
\centering
        \caption{\label{tab:simuhydro} Wind parameters in the hydrodynamic simulation.}
	\renewcommand{\arraystretch}{1.3}
		\begin{tabular}{c c c}
		\hline
		\hline
		Parameters      &  & References\\
        $v_{\infty/WR}$ & 1220\,km/s & \href{http://adsabs.harvard.edu/abs/1992A\%26A...261..503H}{Howarth \& Schmutz (1992)}\\
        $\dot{M}_{WR}$  & 0.8$\times10^{-5}$ M$_{\odot}$/yr & \citet{2002ApJ...566..399M} \\
        $v_{\infty/OB}$ & 2000 km/s   & \citet{2004MNRAS.350..565H}\\
        $\dot{M}_{OB}$        & 0.5$\times10^{-7}$ M$_{\odot}$/yr & \citet{2015PASP..127..428F}\\
        \end{tabular}
\end{table}

\subsection{Spatial sampling of \texttt{RAMSES} quantities}
\label{sec:sampling}
The strong density and pressure gradients occurring in the system impose high refinement along the wind-wind collision interface, as shown in details by \citepads{2011MNRAS.418.2618L, 2012A&A...546A..60L}. The shocked zone is also a region of important cooling, which can only be correctly modeled with a high spatial resolution \citepads{2011CF.....42...44V}. While this is necessary for an accurate representation of the hydrodynamics of the system, this high number of cells is unnecessarily challenging for the radiative effects we focus on in postprocessing.

To retrieve an accurate spatial sampling of the \texttt{RAMSES} outputs and interface them with the radiative transfer code (\S\ref{sec:ramses+radmc}), we interpolate the final output of the simulation over a 3D linear spatial grid. We used the Python-\texttt{RAMSES} interface Pymses\footnote{\url{http://irfu.cea.fr/Projets/PYMSES/intro.html}} to sample the different hydrodynamical quantities, which are the gas density (g/cm$^3$), the pressure (g/cm/s$^2$) and the velocity field (cm/s). The 3D simulation implies a significant increase of elements with the resolution (factor 2$^3=8$) and imposes to choose the sampling parameter N carefully. The primary (non-refined) grid of the \texttt{RAMSES} simulation is sampled over 256$^3$ elements. We choose this value to sample our final grid. The AMR refinement occurs mainly in regions of highest density (i.e., between the two stars and at the immediate surrounding of the WR star); it becomes negligible far from the central core where the dust is expected to form. To ensure that the adopted resolution (N$_r=256$) is sufficient for our purpose, we realize a series of tests using half (128), twice (512) and 4-times (1024) the \texttt{RAMSES} resolution and compute the relevant quantifiable radiative outputs (integrated fluxes, temperatures), which will be described later (Sec. \ref{sec:rnucstudy}). It appears that both the temperatures and fluxes reach a plateau around the adopted resolution (i.e.: equilibrium state) and do not exhibit significant variation after this limit (< 1\%). We note that our results on the geometrical aspect of the wind collision region (Sec. \ref{sec:resulthydro}), are independent of the chosen sampling.

\subsection{Chemical composition and dust formation}
\label{sec:chimicalCompo}

To retrieve the 3d geometrical structure of the dust formed within the WCZ, it is required to convert the gas grid obtained with \texttt{RAMSES} into a usable dusty grid. The exact processes of carbonaceous grain nucleation and subsequent growth are complex and have been studied in various environments such as supernovae ejecta \citepads{2016ApJ...817..134L, 2017A&A...602A.105H}, massive stars \citepads{2000A&A...357..572C} and Asymptotic Giant Branch stars \citepads{2017ApJ...840..117G, 2019A&A...623A.119B}. Dust is the result of a chemical cascade of elemental reactions (radiative combination, molecule-ions reactions, photo-dissociation), initially forming carbon chains of tens of atoms. Those chains then condense other molecular species up to a micrometer size, thus constituting the so-called dust. For the particular case of the WR star environment, \citeads{2000A&A...357..572C} showed that these long carbon chains (typically above six atoms) could only be generated in a dense environment ($\rho>10^{-14}$ g/cm$^3$). Such high densities are present in our simulation but are too close to the stars to allow the dust to survive (above sublimation temperatures). Another way to form carbon dust is to react acetylene molecules (C$_2$H$_2$) to form PAHs (PolyAromatic Hydrocarbons), which are themselves thought to be precursors of amorphous carbon grains \citepads{2014MNRAS.440.1786D}. This second route requires a relatively strong abundance of hydrogen to be effective. Assuming this latter mechanism, we estimate that the most likely zone to form dust efficiently is the mixing zone of the two stellar winds, where the wind of the H-rich main sequence companion will enrich the wind of the C-rich WR star.

To track the relative chemical composition of the fluid inside the simulation, we define two fluid tracers $\chi_{C}$ and $\chi_{H}$. They stand for the relative proportion of the C-rich WR wind and H-rich B star wind in the gas. These two tracers are initialised (to 0 or 1) in each star at each time step, along with the other hydrodynamical quantities. They then evolve freely on the grid. We focus our study on the chemical content of the gas allowing, under certain hypotheses, significant dust formation. We use the local product of the tracers to retrieve the effective mixing of the material in the fluid. The mixing factor is computed as $\zeta = \chi_{C}\chi_{H}$, where $\zeta=0.25$ represents a gas composed equally of H-rich and C-rich wind. This so-called mixing region will trace the WCZ and will make up the ideal locus to form dust efficiently. 

To study the impact of the hydrogen enrichment hypothesis, we compute several values of $\chi_{H}$ from 1 to 40$\%$ logarithmically spaced. We chose a logarithmic sampling to focus on  the turbulent zone we identify at low mixing regime (< 5$\%$, Sect. \ref{sec:resulthydro}). Formally, the corresponding mixing factor $\zeta$ will be set between 0.0099 (1$\%$ of H-rich wind for 99$\%$ C-rich wind) and 0.24 (40$\%$ and 60$\%$ respectively).

We use the linearly sampled grid of the gas density described above (Sec. \ref{sec:sampling}) and the mixing factor $\zeta$ to artificially place the dust in the grid. We consider that the dust is present in all grid elements where the H-rich wind proportion reaches a certain value of $\chi_{H}$ (formally represented by $\zeta$). We then compute the dust density grid by applying a gas-to-dust ratio $\xi_{dust}$.

\begin{figure}
    \centering
    \includegraphics[width=0.98\columnwidth]{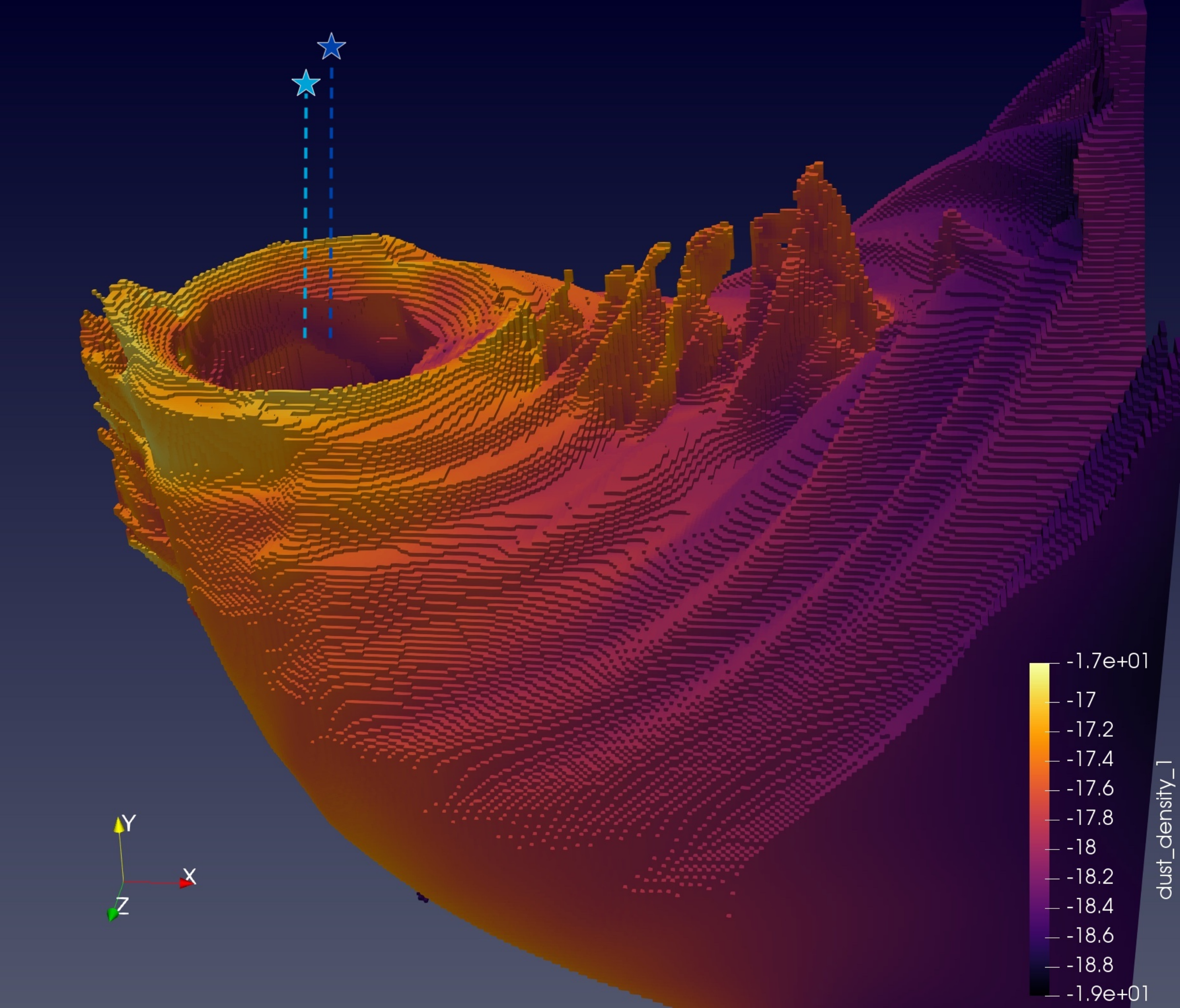}
    \caption{\label{fig:3dview} 3D view of an arbitrary dust density grid ($\chi_{H}=5\%$, $\xi=0.5\%$ and $r_{nuc}=15$ AU). We note the gas/dust plume in the XY plane due to the Kelvin-Helmholtz instabilities (KHI). The positions of both stars are represented (light blue: OB star, blue: WR star).}
\end{figure}

Finally, we use an additional parameter called the nucleation radius $r_{nuc}$ to mimic the dust formation locus. Therefore, we set the dust density to zero in a sphere of radius $r_{nuc}$ around the WR star. This additional parameter allows us to investigate the sublimation temperature of the dust (Sect. \ref{sec:ramses+radmc}). Figure \ref{fig:3dview} shows an example of a 3D representation of the dust grid ($\chi_{H}=5\%$, $\xi=0.5\%$ and $r_{nuc}=15$ AU), where the density is color coded. We also show the positions of both stars (from the \texttt{RAMSES} grid)  for reference.

\section{Radiative transfer modeling}
\label{sec:setupRT}
\subsection{Post-processing of the hydrodynamics simulation}

We use a radiative transfer approach to post-process the dust grid of our \texttt{RAMSES} simulation. The model covers the inner region of the system (56 AU) with a very high spatial resolution (0.007 AU), allowing us to precisely pinpoint the dust nucleation radius. As previously presented, the dust grid is computed using only the gas density grid, a gas-to-dust ratio, a mixing factor and a nucleation radius. Thereby, the aspect of the dust sublimation and survival are not considered \textit{a priori} and need to be inferred from the simulation in an iterative procedure. 

\subsubsection{\texttt{RADMC3d}} 
We used the radiative transfer code \texttt{RADMC3d} \citepads{2012ascl.soft02015D} to determine the dust temperature and the associated dust sublimation radius (Sect. \ref{sec:rnucstudy}).

The robustness and accuracy of the code were validated in the benchmark analysis from \citetads{2009A&A...498..967P}. \texttt{RADMC3d} calculates the temperature distribution of dusty environments, based on a Monte-Carlo method proposed by \citetads{2001ApJ...554..615B}. In practice, a large number (typically a hundreds of million) of packets of photons are launched individually carrying an amount of energy proportional to the total luminosity of the radiation source. Each packet travels through the model dust grid, is absorbed and re-emitted or scattered until it reaches the edge of the grid or disappears (where optical depth is above 30). After launching many photon packets, the equilibrium temperature is then usually reached at each point on the grid. The scattering source function is then calculated for each wavelength using an additional Monte-Carlo iteration. We adopt isotropic scattering, justified by the use of small dust grains (0.1 $\mu m$, Sect. \ref{sec:dustprop}).

\subsubsection{Stellar parameters}
To represent the spectral energy distributions (SEDs) of our system, we used the PoWR\footnote{We used the 2018 version of the WC 06-11 model. The Postdam Wolf-Rayet Models, \url{http://www.astro.physik.uni-potsdam.de/~wrh/PoWR/ powrgrid2.php.}} atmosphere model for the WR star \citepads{2012A&A...540A.144S} and the Kurucz model for the massive companion\footnote{We chose the spectral type B0V (close to B2V) with a log(g)=4.} \citepads{2004astro.ph..5087C}. These advanced atmosphere models enable one to consider the complex phenomena occurring in the photosphere, including emission and absorption features.

Besides the general aspect of the SED, it is necessary to normalise both spectra to compute a realistic total luminosity. Previous estimates appeared to be too coarse due to the important reddening occurring in this galactic region (A$_v$=6.67, \citeads{2020MNRAS.493.1512R}). \citetads{1997MNRAS.290L..59C} reported a total luminosity of 120000 $L_\odot$ with a luminosity ratio (WR:OB) of 0.5 giving $L_{wr}=40000\,L_\odot$, well under the modelled estimations of a single WR star \citepads{2019A&A...621A..92S}. Therefore, we refine this estimate using the high precision magnitudes provided by the Gaia DR2 \citepads{2018A&A...616A...1G}. We use the visual extinction to unredden the three Gaia magnitudes (BP, G and RP). We then normalise the atmosphere models using the luminosity of both components with the Stefan-Boltzmann relation ($L_*=4\pi R_*^2\sigma T^4_*$). We set the effective temperatures to 45000 K and 30000 K for the WR and the B star respectively and use the stellar radii as free parameters. Using the spectroscopic information, we set the intensity ratio in $v$ band (0.5 $\mu$m) to 0.5 \citepads{1997MNRAS.290L..59C}, where the B star is twice as bright as the WR star. We finally adjust the stellar radii to fit the unreddened Gaia magnitudes using a distance of 2.58 kpc \citepads{2018A&A...618A.108S}. We get a total refined luminosity of 270000 $L_\odot$ (Tab. \ref{tab:starsRadmc}).
\begin{table}
\centering
        \caption{\label{tab:starsRadmc} Adopted and refined stellar parameters used to represent the radiation sources.}
	\renewcommand{\arraystretch}{1.3}
		\begin{tabular}{c c c}
		\hline
		\hline
		        & WR star (WC9) & B star (B2V)\\
        T$_{*}$ &  45000 K & 30000 K\\
        R$_{*}$ &  6 R$_{\odot}$ & 13.5 R$_{\odot}$\\
        L$_{*}$ &  140000 L$_{\odot}$ & 130000 L$_{\odot}$\\
        \end{tabular}
\end{table}

\subsubsection{Dust properties}
\label{sec:dustprop}
The dust grain size distribution in WC stars is still debated, ranging from a unique small size of $a = 0.01\,\mu$m \citep{1998MNRAS.295..109Z, 2009MNRAS.395.1749W} to large $a > 0.5$ µm \citep{2001ApJ...550L.207C, 2002ApJ...565L..59M}. In order to proceed in the face of this controversy, we consider three different grain size hypothesis: \textit{small}, \textit{large} and \textit{unique}. Similarly to \citet{2020ApJ...898...74L}, the \textit{small} (\textit{large}) grain size distributions includes grains ranging from $a=0.01-0.1,\mu$m ($0.1-1\,\mu$m) with a number density distribution proportional to $n(a) \propto a^{-3}$. The \textit{unique} grain size hypothesis (i.e., $a = 0.01\,\mu$m) stands for the rapid growth of the dust nuclei limited by the high wind speed velocity \citep{1998MNRAS.295..109Z}.

We assume that the dust grains are composed of purely amorphous carbon grains as suggested by both observations and simulations \citepads{1987A&A...182...91W, 1998MNRAS.295..109Z, 2004MNRAS.350..565H}. The optical constants of the amorphous carbon, represented by the refractive indices (real and imaginary parts), are taken from the database of the Jena University (\url{http://www.astro.uni-jena.de/Laboratory/OCDB/}). The mass absorption and scattering coefficients, $\kappa_{abs}$ and $\kappa_{sca}$, are then computed from these optical constants for the three distribution hypothesis. We consider irregularly shaped grains, approximated by a distribution of hollow spheres (DHS, \citeads{2005A&A...432..909M}).

\subsubsection{Static mesh refinement}
Besides the linear grid used to represent the dust distribution, we perform an additional refinement step before applying the radiative transfer computation. For the Monte-Carlo calculation by \texttt{RADMC3d}, it is essential that the first encountered grid element is optically thin ($\tau<1$). Otherwise, the packet of photons could be caught in this first cell and not contribute to thermalise the overall grid. This effect is demonstrated using \texttt{RADMC3d} on massive thick disks, producing intensity artefacts on the images \citepads{2012ascl.soft02015D}. 

We increase the spatial resolution locally by dividing the selected cell into eight sub-cells (in 3D). These sub-cells can then be divided until reaching the desired optical depth. We applied this refinement on optically thick cells, increasing the level of static mesh refinement (SMR) to get a global optically thin grid. To be conservative and independent of the spectral domain, we perform the refinement using a unique set of absorption and scattering coefficients ($\kappa_{abs}$, $\kappa_{sca}$). We choose the maximal sum of those coefficients ($\kappa_{max}$) compared to the wavelength, and set the limit of the optical depth with:
\begin{equation}
\tau = \rho \times \kappa_{max} \times V_{cells} \geq 1,
\end{equation}
where $\rho$ is the density of a grid element and $V_{cells}$ its volume.

\subsection{Modeling approach}
\label{sec:ramses+radmc}

In order to determine the nucleation radius imposed by the considered grain size distributions (\textit{small}, \textit{large} and \textit{unique}), we perform a series of radiative transfer computations covering a range of nucleation radii from 15 to 35 AU, with a 2.5 AU step. This range of spatial scales is motivated by  previous observational determinations of the nucleation radius \citepads{2008ApJ...675..698T, 2018A&A...618A.108S}. Both report a very similar dust sublimation radius around 12 mas ($\approx31$ AU at the distance of the WR104 system). These measurements were inferred using the observed offset between the peak emission of the dusty spiral structure and the initial point of the fitted Archimedean spiral. Our models cover a broad range of solutions to be sure to include the expected position of this important physical parameter. To examine the impacts of the dust production efficiency, we implement a range of gas-to-dust ratios ($\xi_{dust}=$ 0.1, 1, 3, 5, 7, 10$\%$) and mixing factors ($\chi_H=$ 1, 5, 10, 20$\%$) for each nucleation radii, which represents 648 models in total.

The choice of the dust grain composition is essential to infer the dust nucleation radius imposed by its sublimation temperature. For carbonaceous dust, the sublimation temperature is supposed to be around 2000 K \citepads{2011EP&S...63.1067K}. In the following, we refer as "hot dust" all dust cells presenting an equilibrium temperature above the considered sublimation temperature. If the relative proportion of hot dust becomes significant, we consider the associated radius as too close to be realistic and use the closest non-hot radius as the measured nucleation radius.

\begin{figure}
    \centering
	\includegraphics[width=0.99\columnwidth]{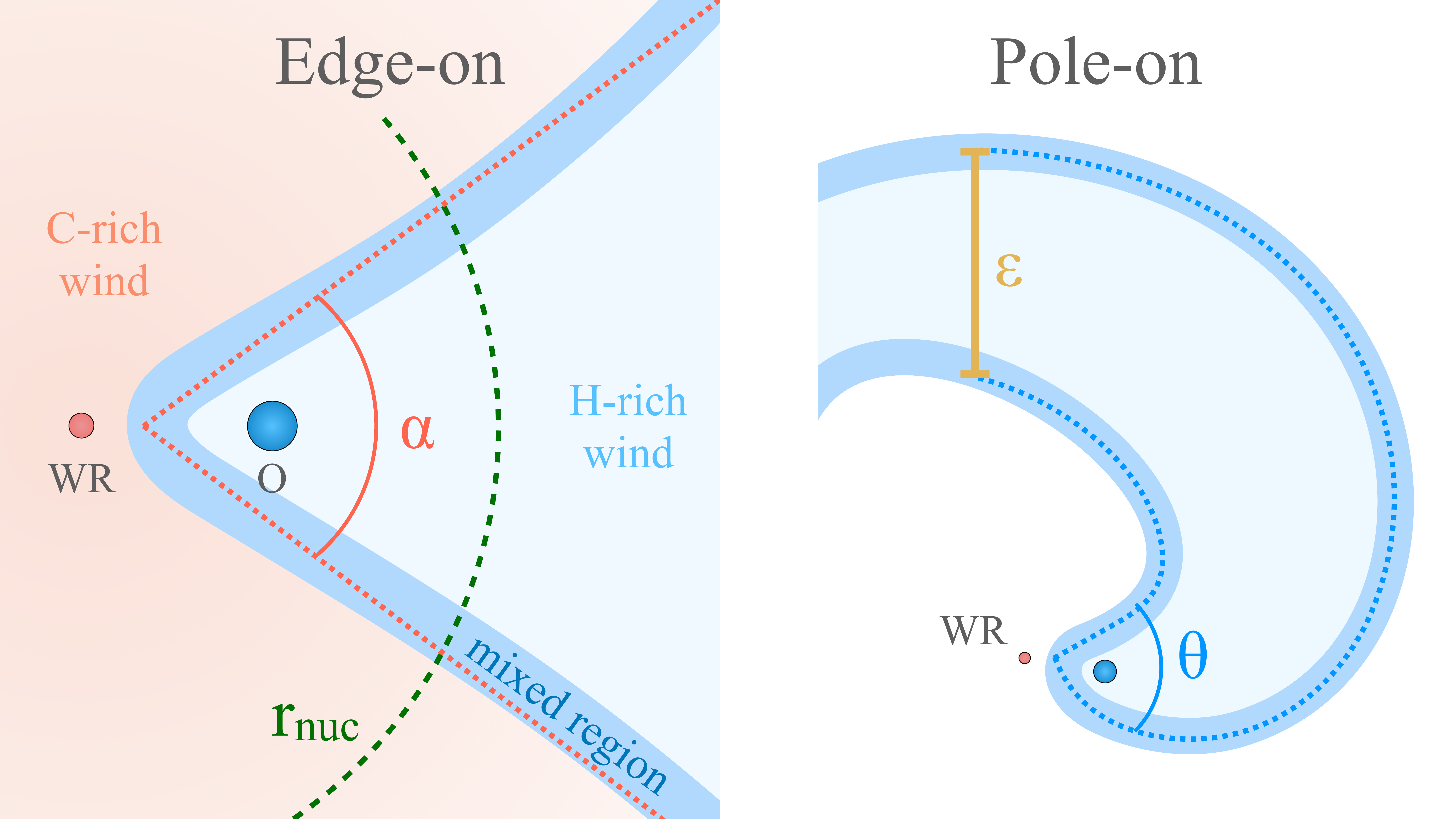}
    \caption{Schematic view of the system presenting the different angles and parameters mentioned in this work. \textbf{Left:} wind-wind collision zone seen from the orbital equator (similar to Fig. \ref{fig:fit_alpha_ex}). \textbf{Right:} Extrapolated spiral seen pole-on.\label{fig:schema_angle}} 
    
\end{figure}


\section{Results and analysis}
\label{sec:results}
\subsection{Parameters arising from mixing}
\label{sec:resulthydro}

In this section, we present the results from the hydrodynamical modeling. Figure \ref{fig:schema_angle} shows the global geometry of the interaction region and the parameters we will be using. The goal is to arrive at a physical intuition for the geometrical parameters obtained by fitting to the observations. 

\subsubsection{Dust opening angle vs. shocked wind-wind interface}

The first parameter we investigate is the full-opening angle $\alpha$. It represents the conical flare of the shocked interface between the two spherical winds. \citetads{1993ApJ...402..271E} states the analytical solution to be proportional to the momentum flux balance of the two stellar winds. In the case of WR~104, the quantities involved give $\alpha\approx40^{\circ}$, yielding one quantitative comparison to confront our numerical estimation with observation.

\begin{figure}
    \centering
    \includegraphics[width=\columnwidth]{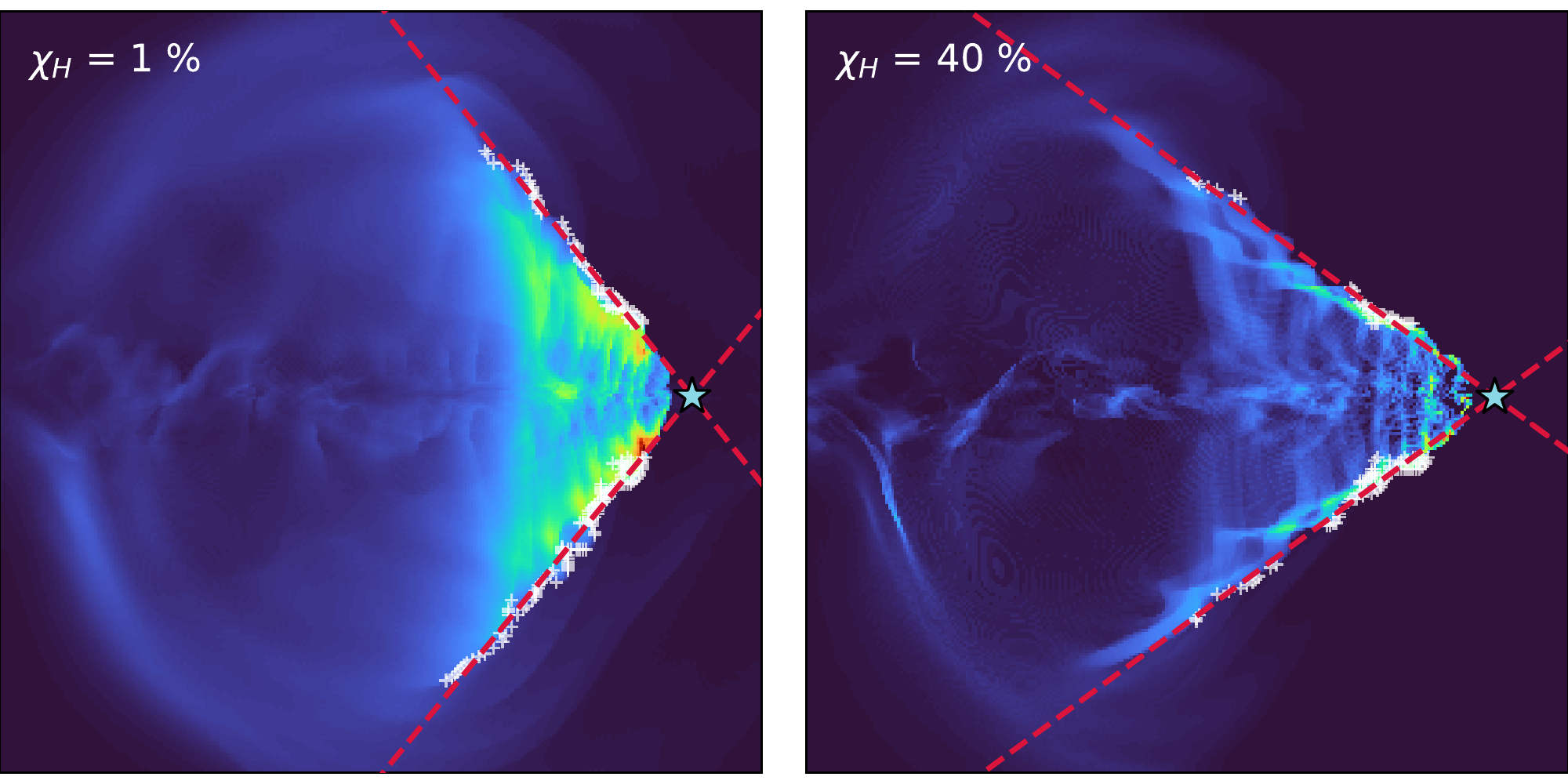}
    \caption{\label{fig:fit_alpha_ex} 
    Example fits to the opening angle on the projected edge-on density grid. \textbf{Left:} Low mixing regime, \textbf{Right:} High mixing regime. The automatically detected edges are represented by white plus symbols, the position of the WR star as blue star and the fit as red dashed lines.}
\end{figure}

We derive the opening angle for each mixing value from 0.0099 to 0.24, representing the hydrogen enrichment of the gas (Sect. \ref{sec:chimicalCompo}). We use a Laplacian filter applied on the edge-on projected grid to fit both sides of the cone. To automatically detect the edges, we add distance and density conditions to avoid strong deformation of the central region (Fig. \ref{fig:fit_alpha_ex}). We use a Levenberg-Marquardt minimisation to determine both the values and the statistical uncertainties of $\alpha$, recovering an average precision of $\sigma_\alpha=0.6^\circ$. The adjusted parameters are plotted in Figure \ref{fig:result_hydro_mix}, illustrating the two most extreme values of H-enrichment ($\chi_{H}=1\%$ and 40$\%$) under consideration. 

We identify two different regimes, hereafter referred to as the low and high mixing regime. Both regimes can be remarkably well reproduced by a linear model, allowing us to estimate the opening angle for any mixing factor. We set two $\chi_{H}$ ranges ($1<\chi_{H, 1}<4\%$, $5<\chi_{H, 2}<40\%$) to fit the slope and the origin point (Tab. \ref{tab:result_coeff}). 

In the low-mixing regime (regime 1), the opening angle grows rapidly with the mixing factor. As the desired proportion of hydrogen is reduced, nucleation gets closer to the primary shock of the spiral. We can, therefore, push the hydrogen proportion to the limit of $\chi_{H}=0.001\%$ to be as close as possible to the fully-carbonaceous region of the shock. Doing so yields $\alpha=122\pm1^\circ$, three times larger than the \citetads{1993ApJ...402..271E} approximation. \citetads{2016MNRAS.460.3975H} already reported such discrepancy in the comparable system of WR98a. Similarly, this high amplification of the opening angle appears to be caused by the effective cooling occurring in the shocked interface.

In the high-mixing regime (regime 2), the increase relative to $\chi_{H}$ is smoother and appears to be twice as big as the reported values of 35/40$^\circ$ obtained with high angular resolution observations \citepads{2008ApJ...675..698T, 2018A&A...618A.108S}. \citet{2004MNRAS.350..565H} reported that the opening-angle should not exceed 80 degrees, constrained by masking interferometry data. These limiting values are still debated today due to insufficient angular resolution and image fidelity, so that we cannot entirely rule out the more extensive range of opening-angles encompassed in this work.

\subsubsection{Internal structure of the spiral}
\begin{figure}
    \centering
	\includegraphics[width=\columnwidth]{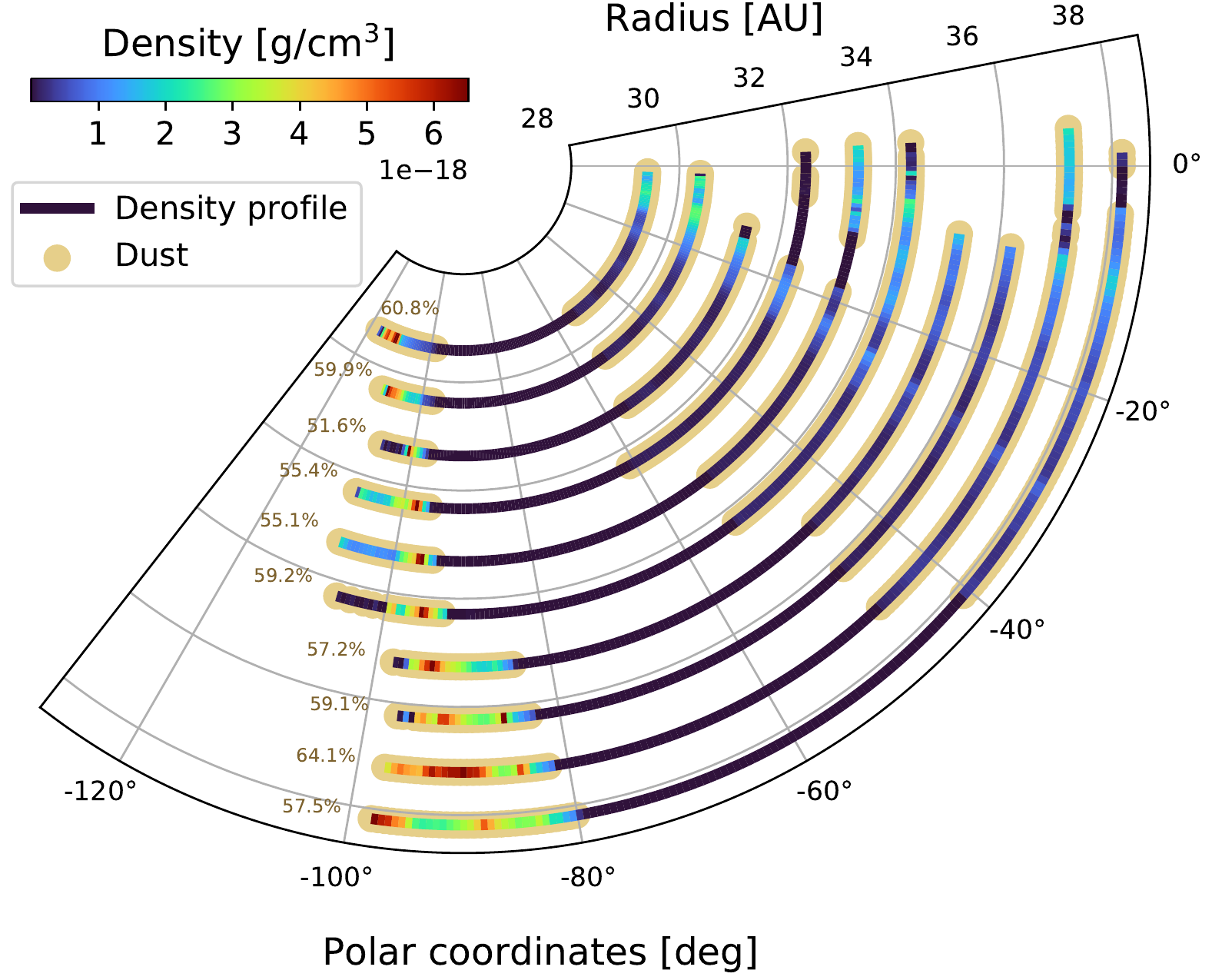}
    \caption{Density profiles in the middle-plane at different radii. The binary is located at the origin's axis (R = 0 AU) not represented in this plot. The left/thinner side is the leading edge of the spiral (same orientation of Fig. \ref{fig:velocity_map}). The detected dust is represented as beige dots. We show the measured fraction $\omega$ on the left of each profile.}
    \label{fig:inter_struct}
\end{figure}

The mixed region considered to produce the dust forms an empty conical structure, represented by a filling factor $\omega$. This new geometrical parameter stands for the relative wall thickness of the spiral and will play a central role in the brightness distribution at a larger scale. To investigate this parameter, we explore the mid-plane (Y--X plane, Fig. \ref{fig:3dview}) grid of the dust model to get a reasonable estimate of $\omega$. The internal structure appears to be highly heckled and shows a large variety of thickness. Figure \ref{fig:inter_struct} show the internal density profile computed at different radii in the middle-plan of the spiral. We determine the relative proportion of the dust included within both the leading and trailing walls.

For each mixing regime, we compute the average and the standard deviation of the radially dependent $\omega$. The considered uncertainties let to keep track of the radial distribution of $\omega$. Anew, two regimes can be recognised and appear to be linearly dependent of $\chi_H$. We note that the two regimes are localised in the same region as for the opening-angle, suggesting a similar physical cause. Nonetheless, the transition around $\chi_H=5\%$ does not exhibit a significant slope difference (Fig. \ref{fig:result_hydro_mix}). The fitted linear coefficients $a$ and $b$ are presented in the Table \ref{tab:result_coeff}.

\subsubsection{Face-on angle $\theta$ and arm width $\epsilon$}

In this section, we describe our investigation concerning the spiral arm width ($\epsilon$ in the following, Fig. \ref{fig:schema_angle}). Despite the relatively limited simulated field size of our simulation, we were able to arrive at an estimate of this critical parameter using the face-on density image. 

The first step consists of measuring the angular shift between the two sides of the spiral seen face-on. Both leading and trailing-edges can be described by an Archimedean spiral, and the rotation between those two geometric spirals corresponds to the measurement of the face-on angle $\theta$. We note that this measurement is supposed to be identical to the edge-on opening angle ($\alpha$), but appear be slightly different due to projection effects. Therefore, we address this discrepancy by using an alternative method to measure the face-on opening angle $\theta$, distinct from $\alpha$ reported before (Fig. \ref{fig:schema_angle}). We measure $\theta$ using the polar density profiles at different radii (similar to Fig. \ref{fig:inter_struct} but on the summed face-on image). We used ten profiles to determine the average value and its dispersion. Thus, $\theta$ is simply given by the angular distance between the two extremities of the profile above a threshold density value (i.e.: the position of the archimedean spiral associated with each side). 

We then extrapolate the spiral shape at large scale using an Archimedean spiral model to represent both edges of the spiral separated by the measured angle $\theta$. To do so, we need to define the spiral step $S$ of the Archimedean model. We explore different values to fit both the edges and the centre of the simulation. We determined that the best match is obtained with $S=170$ AU, fully consistent with the observations ($170\pm8$ AU, \citeads{2018A&A...618A.108S}).

\begin{figure}
    \centering
	\includegraphics[width=0.99\columnwidth]{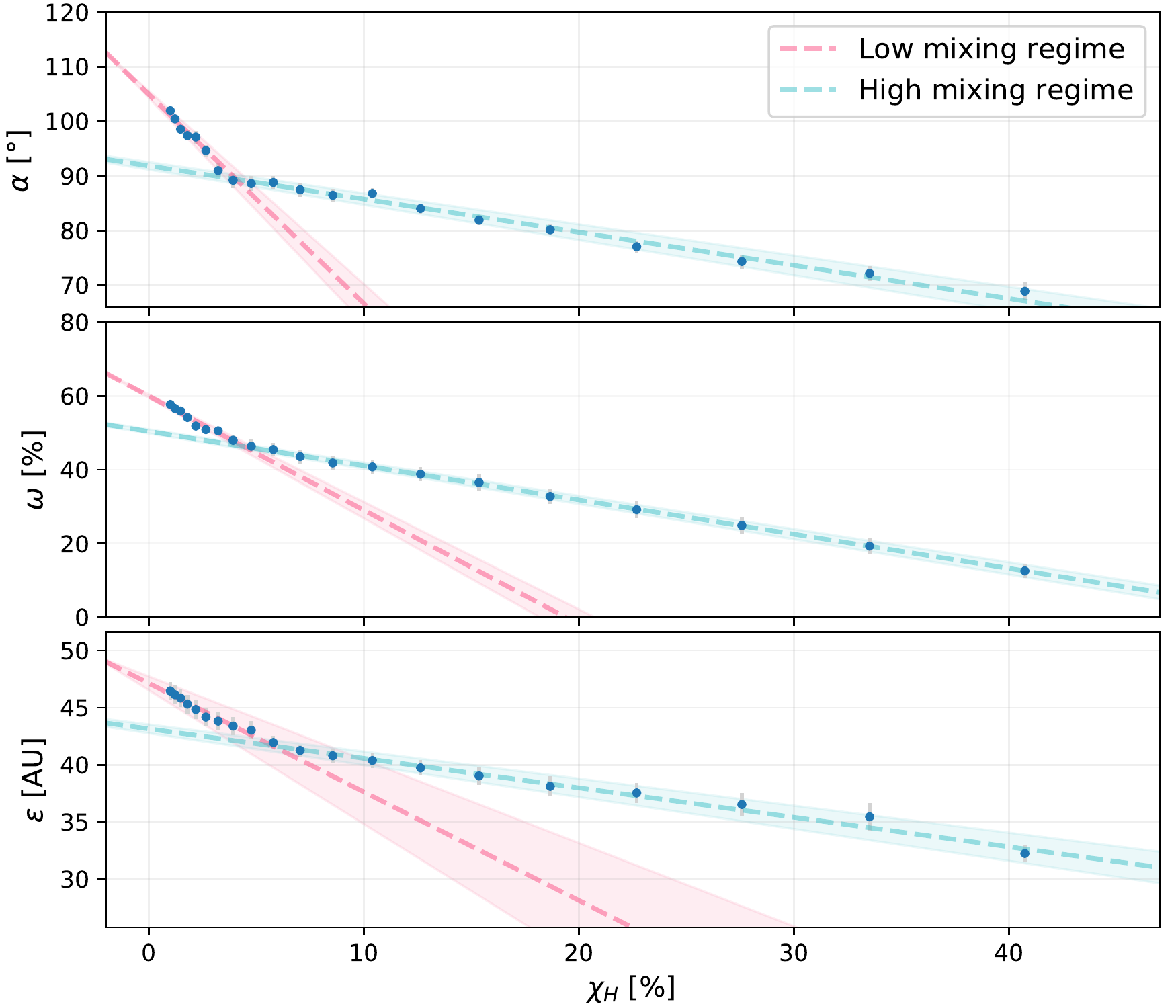}
    \caption{Evolution of the pinwheel parameters according to the hydrogen proportion ($\chi_H$). The presented parameters are the opening angle $\alpha$, the filling factor $\omega$ and the spiral width $\epsilon$. The blue dots represent the measured values. The two regimes we identify are represented in red (low mixing) and cyan (high mixing). Both linear fit parameters are shown in Tab. \ref{tab:result_coeff}.}
    \label{fig:result_hydro_mix}
\end{figure}

The arm width $\epsilon$ can then be measured as the linear distance between the two spiral models at a given azimuth (sufficiently far from the origin; see Fig. \ref{fig:rnuc_study}). The results are presented in Figure \ref{fig:result_hydro_mix}. We fit the dependency of mixing using the same linear approximation method for both mixing regime (Tab. \ref{tab:result_coeff}).

\begin{table}
\centering
        \caption{\label{tab:result_coeff} Fitted linear model used to represent the geometrical parameters. $a$ and $b$ refer to the slope and the origin point respectively (e.g.: $\alpha=a\chi_H+b$)}
	\renewcommand{\arraystretch}{1.3}
		\begin{tabular}{c c c}
		\hline
		\hline
		Parameters & Regime 1 & Regime 2\\
                   &  $a=-3.83\pm0.29$ & $a=-0.61\pm0.04$\\
        Full-opening angle $\alpha$ &  $b=104.9\pm0.6^\circ$ & $b=91.9\pm0.7^\circ$\\
                   &  $\chi^2_r=1.06$ & $\chi^2_r=0.59$\\
                   &  $a=-3.09\pm0.18$ & $a=-0.93\pm0.03$\\
        Filling factor $\omega$ &  $b=60.0\pm0.4\%$ & $b=50.4\pm0.6\%$\\
                   &  $\chi^2_r=2.64$ & $\chi^2_r=0.12$\\
                   &  $a=-0.95\pm0.22$ & $a=-0.26\pm0.02$\\
        Arm width $\epsilon$ &  $b=47.1\pm0.6$ AU & $b=43.1\pm0.4$\\
                   &  $\chi^2_r=0.13$ & $\chi^2_r=0.19$\\
        \end{tabular}
\end{table}

\subsection{Dust vs. gas velocity}
\label{sec:dustvelocity}

The dust velocity is typically attributed to the dominant terminal wind speed of the Wolf-Rayet star \citepads{1999Natur.398..487T, 2008ApJ...675..698T}. However, this strong hypothesis is still debated in the literature and arguments exist to attribute the dust expansion to the weaker companion wind \citepads{2008MNRAS.388.1047P, 2004MNRAS.350..565H}. This assumption is particularly relevant to constrain the astrometric distance to the system. Indeed, the spiral step associated with the orbital period allows computation of the apparent angular velocity, which can be combined with the physical speed of the dust plume to yield the distance \citep{2018A&A...618A.108S}. In this section, we use the velocity map rendered by \texttt{RAMSES} code to investigate this hypothesis.

\begin{figure*}
     \centering
     \begin{subfigure}[b]{0.47\textwidth}
         \centering
         \includegraphics[width=\textwidth]{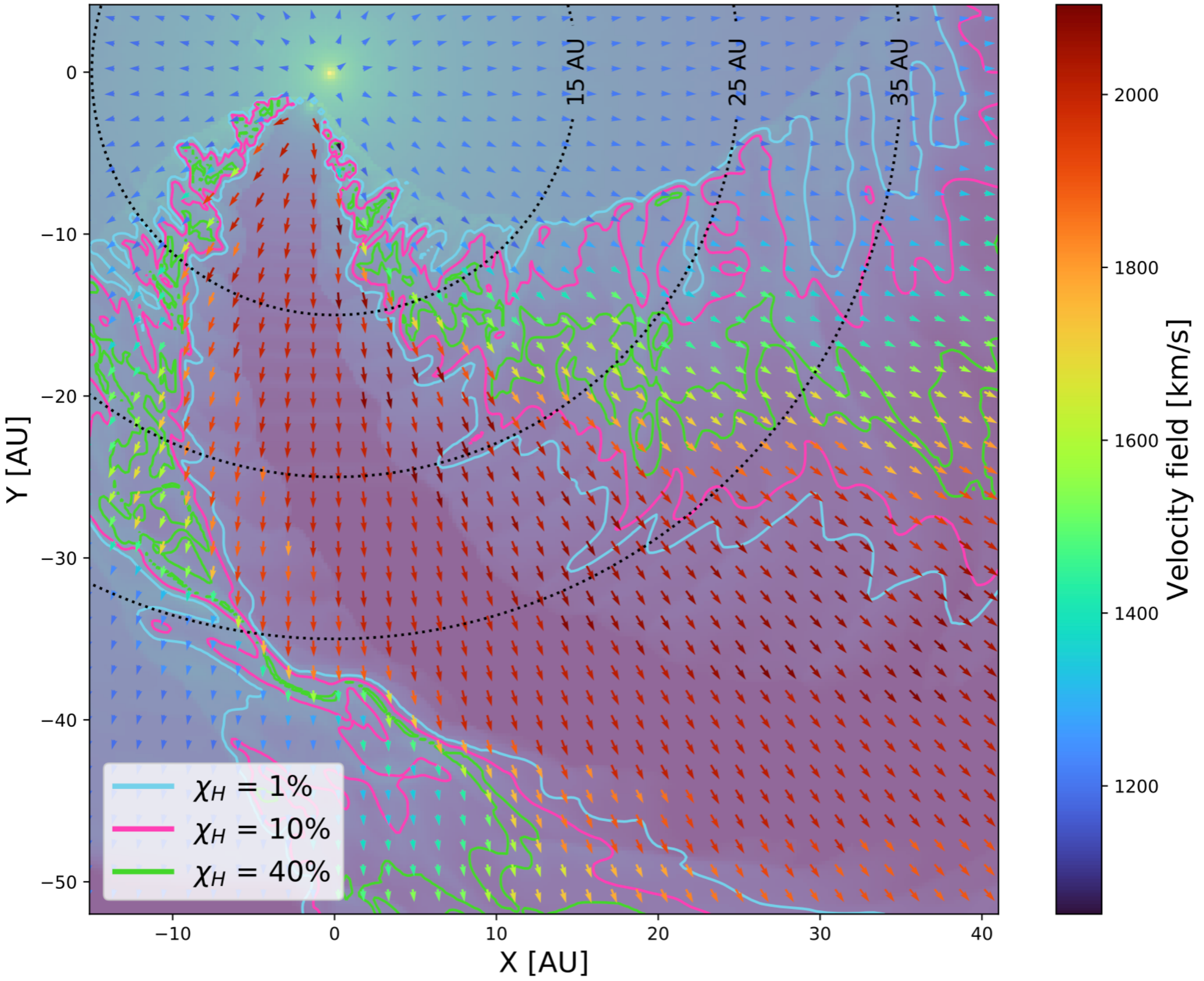}
         \caption{Velocity map in the orbital plane of the binary.}
         \label{fig:velocity_map}
     \end{subfigure}
     \hfill
     \begin{subfigure}[b]{0.5\textwidth}
         \centering
         \includegraphics[width=\textwidth]{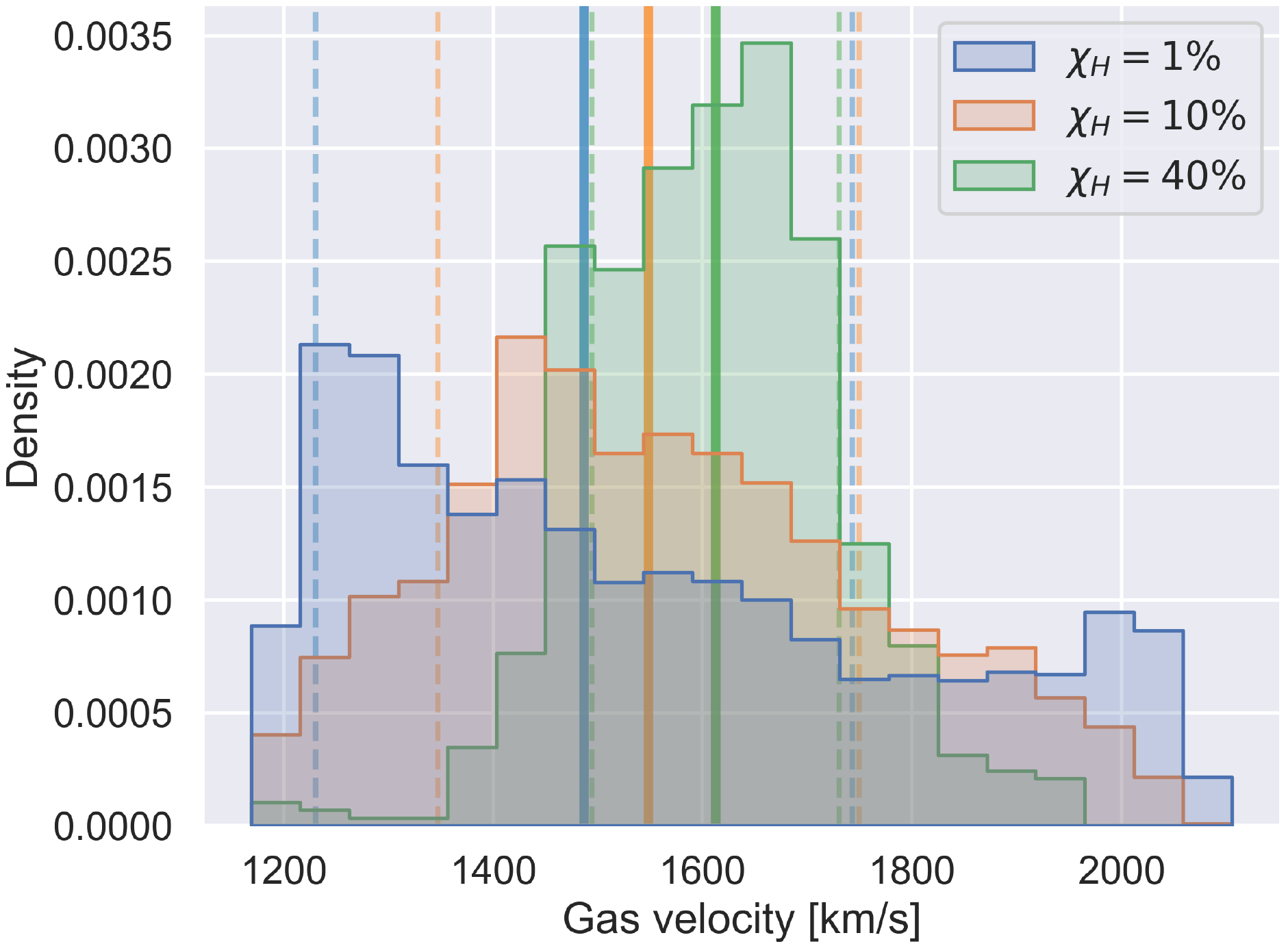}
         \caption{Normalized distribution of the gas velocity.}
         \label{fig:velocity_hist}
     \end{subfigure}
     \hfill
        \caption{2d map and associated distribution of the gas velocity along the orbital plane of the binary. \textbf{a)} The underlying color-scale layer shows the density map (low opacity). The quiver plot (overlaid arrows) represents the velocity field, in direction and amplitude and is coloured according to the absolute wind speed (km/s). The coloured contours represent different hydrogen enrichments (1, 10 and 40$\%$). Three possible dust nucleation radii are illustrated as black dashed lines. \textbf{b)} The three colors represent different hydrogen enrichments (blue: 1$\%$, orange: 10$\%$ and green: 40$\%$). The vertical lines are the average values.}
        \label{fig:velocity}
\end{figure*}

Figure \ref{fig:velocity_map} shows the gas velocity in the binary orbital plane. The ``quiver plot'' illustrates the amplitude and direction of the velocity field whereas the color indicates the value. In addition, we show the gas density map as the lower layer (with low opacity) as a reference of location. For our purpose, we display as colored contours the different layer of H-enrichment (cyan: 1$\%$, pink: 10$\%$ and green: 40$\%$). 

We measure the velocity at the different expected locations of dust production (i.e.: distance from the WR star) and different H-enrichments. We report a similar velocity field across the different distances we tested (as showed as dashed in Fig. \ref{fig:velocity_map}). The terminal wind speed is, as expected, already reached at the closest distance of 15 AU. Therefore, the conclusion focused on the hydrogen dependency is uncorrelated to the nucleation radius (as reported in Sect. \ref{sec:rnucstudy}).

We find a high discrepancy between the two most extreme scenarii (1 and 40$\%$ of H-enrichment). For the high mixing regime, the gas velocity is perfectly balanced between the two stellar components being around $1612 \pm 117$ km/s (Tab. \ref{tab:velocity_gas}). The velocity is relatively constant within the highly mixed thin layer (green line in Fig. \ref{fig:velocity_map}). We measure a similar velocity field on both leading and trailing arms, with however a very strong gradient on the inner edge of the leading arm. The overall velocity field appears to be strongly correlated to the mixing ratio exhibiting similar gradient on both arms.

The low mixing regime exhibits highly differential velocity values across the considered layer (inside the cyan line, Fig. \ref{fig:velocity_map}). The gas is dominated by the O-star component in the inner edge (1800-2000 km/s), whereas it is dominated by the WR-wind on the outer edge (1000-1200 km/s). We report a median velocity of $1486 \pm 256$ km/s, where the indicated uncertainty stands for the dispersion of the distribution (Tab. \ref{tab:velocity_gas}). For comparison, we present the normalized distribution of the velocities between 15 and 40 AU for three $\chi_H$ (1, 10, 40$\%$). The relative flat distribution of the low-mixing scenario exhibits a large gas speed dispersion across the spiral walls, whereas it is peaked around 1600 km/s for the high--mixing one (Fig. \ref{fig:velocity_hist}). The resulting dust expansion speed submitted to this differential velocity field between its edges could results in over-densities at larger scales. 

\begin{table}
\centering
        \caption{\label{tab:velocity_gas} Velocities of the gas as function of the hydrogen enrichment $\chi_H$.}
	\renewcommand{\arraystretch}{1.3}
		\begin{tabular}{c c c c}
		\hline
		\hline
		 $\chi_H$ [$\%$] & Medians [km/s] & Dispersion [km/s] & Extremes [km/s]\\
        1  & 1513 & 257 & 1069--2105\\
        10  &  1548 & 204 & 1069--2080\\
        40  &  1616 & 129 & 1168--1960\\
        \end{tabular}
\end{table}

\subsection{Dust nucleation radius}
\label{sec:rnucstudy}

\begin{figure*}
  \sbox0{\begin{tabular}{@{}c@{}}
    \includegraphics[width=1.\paperwidth]{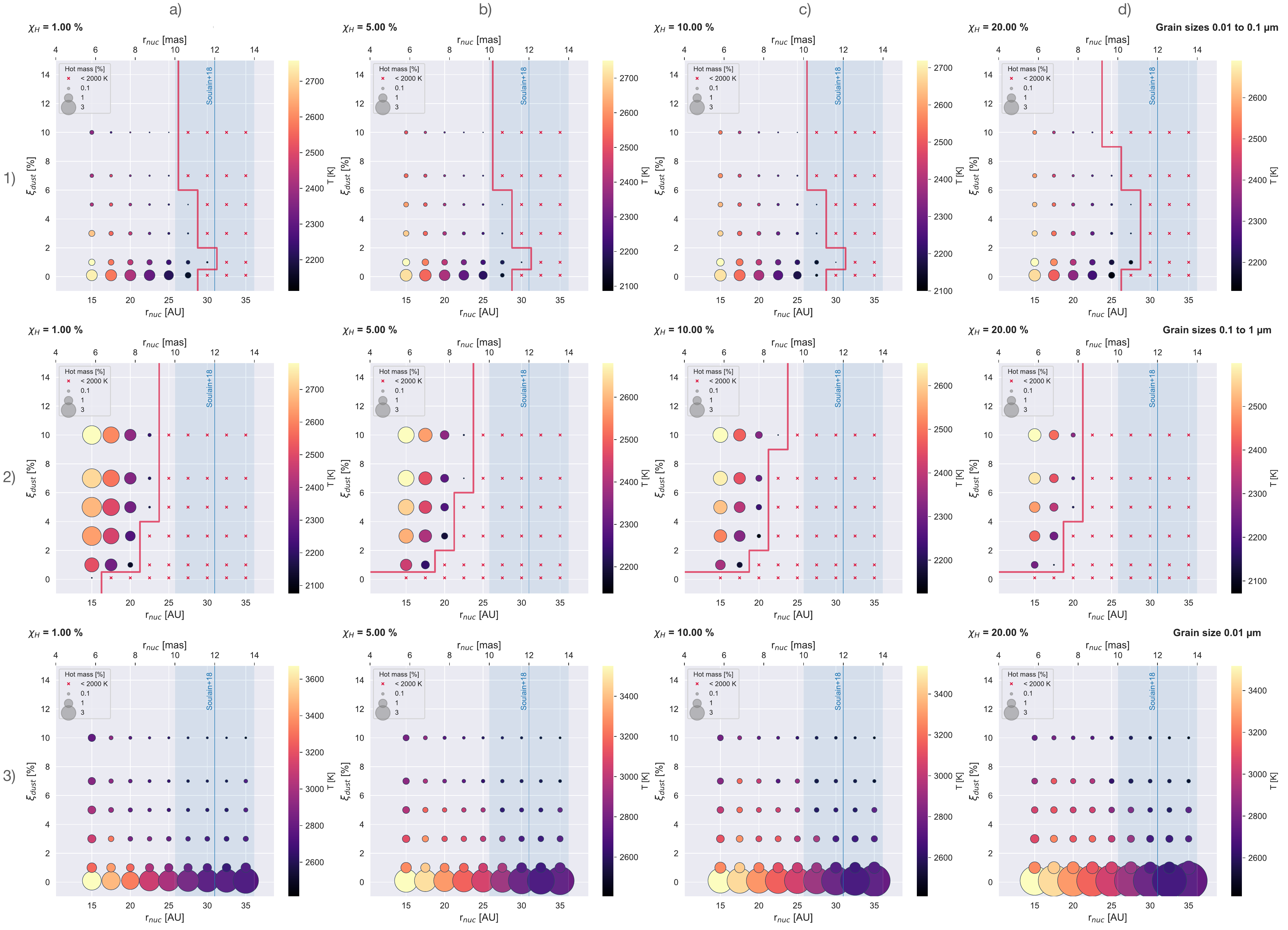}
  \end{tabular}}%
  \rotatebox{90}{\begin{minipage}[c][\textwidth][c]{\wd0}
    \usebox0
    \caption{\label{fig:rnuc_study}Maximum temperature (color gradient) and hot cell fraction (scatter size) compared to the the nucleation radius and the gas-to-dust ratio. Each dot corresponds to one of the 648 radiative transfer models computed in this work. The red crosses represent the case where the dust temperature is bellow the sublimation limits of 2000K. \textbf{From left to right:} mixing values from 1 to 20$\%$. \textbf{Top:} \textit{small} grain size hypothesis. \textbf{Middle:} \textit{large} grain size hypothesis. \textbf{Bottom:} \textit{unique} grain size hypothesis. The red line corresponds to the predicted nucleation radius determined in this work. The observational estimate is shown in blue for comparison \citepads{2018A&A...618A.108S}.}
  \end{minipage}}
\end{figure*}

In this section, we discuss the different constraints we have on the nucleation radius. Figure \ref{fig:rnuc_study} shows the relative proportion of hot cells compared to the nucleation radius and the gas-to-dust ratio for different  dust composition hypotheses (in columns, a to d) and grain sizes (in lines, 1 to 3). For the specific combination of H-enrichment ($\xi_H=5\%$) and grain sizes (\textit{small} -- see \textit{b1} in Fig. \ref{fig:rnuc_study}), the expected nucleation radius appears to be close to 30 AU for the overall the range of $\xi_{dust}$. We note the same behaviour for the other values H-enrichment, with an expected nucleation radius around 30-32.5 AU (Fig. \ref{fig:rnuc_study}).

For a specific nucleation radius, an increase in mass reduces the relative proportion of hot dust while decreasing the maximum temperature. We note the nucleation radius seems to be independent of the dust mass, in the range tested by our models. Regarding the maximum temperature, our models show quite comparable results for the enrichment cases we explored. They differ from 2100 K to 2800 K for the most extreme scenario, close to the star (15 AU) and with the lowest enrichment (1 $\%$).

For the \textit{large} grains scenario, the figure \ref{fig:rnuc_study} shows a nucleation radius closer to the star, between 20 AU and 22.5 AU. A nucleation radius this close to the central source is very unlikely and can be excluded by observational constraints. The measured radius is expected to be even closer for the smallest gas-to-dust ratio we tested (0.1 $\%$) and cannot be constrained by our model. The maximum temperatures are similar to the previous case but can reach higher values, up 3000 K, while affecting a larger fraction of mass (up to 5$\%$, against 2$\%$ for the \textit{small} scenario).

Finally, the \textit{unique} grains size scenario does not allow for any dust condensation in the parameters explored in Figure~\ref{fig:rnuc_study}. The four  enrichment cases are always above the sublimation temperature of 2000 K, even for the tiniest mass fraction. Dust never condenses in such a case but could condensate farther away (beyond the boundaries explored in this study). Additionally, the maximum temperatures are much higher than the two previous scenarii, standing between 2500 K and 3700 K. The fraction of hot dust is also higher, up to 10$\%$. Interestingly, the fraction of hot dust increases with the H-enrichment, contrary to the two previous scenarios.

\section{Discussion}
\label{sec:discussion}

Despite only simulating a finite circumstellar volume around the WR~104 binary dedicated to the innermost regions, we have recovered several insights into the system's properties.

\subsection{Proxy of dust nucleation process}

Based on purely geometrical arguments, our \texttt{RAMSES} simulation suggests the presence of two distinct nucleation regimes depending on the hydrogen enrichment of the gas. All of the three geometrical parameters explored by our study (i.e. the opening angle $\alpha$, the filling-factor $\omega$ and the spiral arm width $\epsilon$) appear to be linearly correlated with the hydrogen enrichment parameter ($\chi_H$). The first regime, which we called the "low mixing regime", corresponds to a relatively fast decrease of the parameter with respect to the $\chi_H$. For instance, the opening angle changes from 105 to 90 degrees on the first 5$\%$ of hydrogen variation, while falling to 70 degrees at the maximum considered enrichment. Such discrepancy, between low and high regimes, appears less important for the two other quantities, being less affected by the geometry of the external spiral walls. For the most extreme case (i.e.: $\chi_H=1\%$), the dust locus reaches the interface between the gaseous spiral cone and the WR free-wind dominated region. Such regions can be subject to various hydrodynamic instabilities. They are responsible for the wavy aspect of the shocked region in the 3d grid in Fig. \ref{fig:3dview}. The velocity difference between the winds leads to the development of the Kelvin–Helmholtz instability (KHI). As the velocity gradient at the interface is small, the growth of the instability remains small and the instability is linear \citep{2011MNRAS.418.2618L}. Such unsteady interface physics could explain the reported difference between both regimes.

The two regimes we identified therefore represent a good proxy for the nucleation processes, that can be compared with observational constraints. All previous studies focused on pinwheel nebulae \citep{1999ApJ...525L..97M, 2008ApJ...675..698T, 2018A&A...618A.108S, 2020ApJ...900..190L} report various estimates of morphological parameters (e.g.: opening angle from 40 $\deg$ to 120 $\deg$). With our study, we propose to correlate such geometrical measurements with underlying physical properties represented by the hydrogen enrichment. In particular the opening angle seems to exhibit large variation depending of the H-enrichment and could therefore represent a sensitive diagnostic of the nucleation processes. 

Additionally, we report a new estimate of the value of the filling factor. This important parameter is widely considered in the literature but without clear definition. \citet{2008ApJ...675..698T} and \citet{2018A&A...618A.108S} reported an empty spiral structure bounded by an infinitely thin surface sheet: an approximation that entirely sidesteps any filling proportion arguments. In our study, we report values ranging from 20 to 60$\%$ that may affect significantly the spiral arm's intensity profile. 

We address the question of the spiral arm width by analysing the projected opening angle at a larger scale. The arm width is usually derived from the opening angle of the wind collision zone in a purely geometrical approach. In this study, we report a more realistic determination by considering the 3d aspect of the deformed cone. The opening angle seemed slightly different between the orbital plane view and the orthogonal polar plane. The reported width corresponds to the orbital plane, where the spiral is seen face-on. Figure \ref{fig:result_hydro_mix} reveals the imperfect correlation between the opening angle of the shock (not deformed by the orbital motion) and the large scale spiral arm width. Therefore, observers should assess the typical correlation between those two fundamental aspects and carefully take account of the degree of projection in deriving their model. The linear correlation found in this work can accordingly be used as a calibration hypothesis to retrieve the accurate opening angle from the spiral width determination and the other way around.

\subsection{Dust nature and nucleation temperature}
\label{sec:temperature}

The radiative transfer modeling allows us to get an independent measurement of the nucleation radius. This crucial parameter appears to not be related with the dust-to-gas ratio (i.e. the density), the latter only affects the proportion of "hot" dust in the simulation. Considering the observational estimates of the nucleation radius (10-14 mas, \citepads{ 2008ApJ...675..698T, 2018A&A...618A.108S}, we are able to draw conclusions about the nature of the dust grains, which are compatible with a small grain size distribution (0.01-0.1 µm). When considering a unique grain size (0.01 µm), the absence of nucleation lets us rule out the hypothesis of no evolution \citep{1998MNRAS.295..109Z}. We note that our favoured small grain size distribution appears to be in contradiction with the work of \citet{2020ApJ...898...74L} which focused on the analysis of the Spectral Energy Distribution (SED) on a sample of WC stars.

The nature of the dust around WR stars appears different from the interstellar dust \citep{1994ApJ...422..164K}, Young Stellar Object \citep{2010Natur.466..339K} or the Asymptotic Giant Branch (AGB) dust \citep{2006A&A...447..553F, 2020A&A...641A.103V}. Regarding the dust species formed in oxygen-rich environments (eg. AGBs), the most stable compounds are silicates, alumina dust (Al$_2$O$_3$), and solid iron. In the case of the WR stars, C-rich species are favoured by formation pathways through the PAH reaction and should have a limited size distribution due to their rapid growth and subsequent rapid (1200--1600 km/s) injection into the ISM. 

The carbonaceous dust could have a larger than predicted impact on the overall dust amount of galaxies \citep{2009MNRAS.396..918M, 2012ApJ...748...40B, 2020ApJ...898...74L}; redistributing the energetic stellar budget in a different manner (by way of absorption efficiency and diffusion). In the context of galactic chemistry, the presence of moderate size grains (0.01--0.1 µm) could favor the transport of Alumina isotopes in Solar-type nebulae \citep{2017ApJ...851..147D}. Depending on the total contribution of the dust of WR origin in the ISM, bigger and more resistant carbonaceous grains should have a non-negligible impact on the destruction timescale of interstellar dust. More specifically, larger carbon grains could survive to the supernova and post-supernova shock \citep{2006ApJ...648..435N, 2016ApJ...817..134L}, following the WR stage of massive stars and could therefore contribute significantly to the dust enrichment of galaxies. Especially in the early Universe, where the total amount of massive stars was significantly higher than today \citep{2003MNRAS.343..427M}.

\subsection{Differential velocity vs. clumpy spiral plume}

We saw that depending on the dust nucleation pathways, the newly formed dusty grains can be ejected at different speeds (Fig. \ref{fig:velocity_hist}). For the most extreme scenario, the velocity could exhibit a strong gradient ranging from 1200 to 2000 km/s. If some dusty structures are faster at their birth origins, some overlap could therefore appear after some orbital revolutions. Such over-densities are hardly distinguishable from the projection effect showing brighter spiral edges \citep{2016MNRAS.460.3975H} and should be explored at larger scales. Recent simulation studies, such as \citet{2022arXiv220407397E}, report a very clumpy dusty structure that could be addressed by such differential velocity origins.

\subsection{Leakage of dust and instabilities}
\label{sec:instabilities}

The external walls of the spiral wind interface appear to be dominated by the Kelvin-Helmholtz instability (KHI). The effect of the instability can be seen in figure \ref{fig:3dview} as large plumes or `fingers' escaping the cone of the spiral. These fingers have been previously reported by \citet{2015MNRAS.449.3780M}, using a SPH simulation of the Eta Carinae system, or by \citet{2012A&A...546A..60L} using the \texttt{RAMSES} code on a variety of wind parameters. In theory, such instabilities should be able to carry dust particles out of the confined gaseous spiral and enrich the nebulae all around the binary system. We inspect the previous stage of our simulation to retrieve the previous positions of these plumes. Our simulation spans just enough time to retrieve the spiral deformation due to the orbit, however past stages are not sufficiently advanced to be used at a confident level. Nevertheless, these structures are present in the previous output but appear smaller and closer to the star. Therefore, they could fill-up the entire domain of the simulation after few orbits. This phenomena could amplify the dust nucleation process occurring within the pure WR wind, as recently reported in the hydrodynamical simulation of various colliding wind binaries \citep{2022arXiv220407397E}. Such global structure has been reported around various colliding binaries using long-baseline interferometry techniques \citep{2009A&A...506L..49M, 2016SPIE.9907E..3RS}. Such studies indicate the presence of an extended dust "halo" responsible for very low visibility measurements recovered by interferometry. We are actively investigating whether such a halo can be attributed to a fully resolved circumstellar shell, however such results lie beyond the scope of the present paper and we defer discussion to a later publication. 

\section{Conclusions}
\label{sec:conclusion}

Inspired by observational findings for the WR 104 system, we have used hydrodynamical
simulations and radiative-transfer post-processing to study the
properties of the dust nucleation process occurring around massive evolved binary stars. We investigate the 3D structure of the colliding wind binary in the innermost region where dust is expected to form. Below we summarize our most important results.

(1) We report a direct correlation between the observable parameters governing the form of the observed spiral plume (opening-angle, spiral width and filling-factor) and the chemical composition of the gas; represented by the hydrogen-enrichment of the O-type star companion. 

(2) The velocity field shows a differential structure that could challenge the standard assumption about the terminal wind speed of such systems. Depending of the dust nucleation locus (i.e. the H-enrichment), the projected velocity of the dust can present a strong gradient (1200-2000 km/s) or a more balanced speed (1600 km/s). Such discrepancies should be carefully considered when comparing the spectroscopic measurement of the stellar wind-speed and the projected dust speed.

(3) The nucleation radius stands around 25-30 AU if we consider the small grain size distribution (0.01--0.1 µm), in  close agreement with the observational constrains. The chemical nature of the dust around WR stars diverges from that of other commonly-encountered dust producing systems in the galaxy; a finding that could have a non-negligible impact on galactic dust composition.

(4) High levels of instability at the shock interface creates finger structures that, in turn, could facilitate escape or material through the collimated gaseous wall. This may carry dust outside the spiral, and could precipitate the presence of dusty "halo" around CWB systems for which observational support already exists from interferometry.

We demonstrate in this paper how the hydrodynamical simulation is a critical tool to understand the physics of these complex astrophysical systems. The realistic 3D view obtained with \texttt{RAMSES} code can then be post-processed by a radiative transfer code to extract the valuable quantities. The colliding wind binary systems represent a challenging science application, requiring diverse timescales, spatial scales and the inclusion of a multitude of physical phenomena (eg. radiative cooling, instabilities, dust opacity). The  WR~104 system appears even more complicated than expected with the reported differential velocities, internal structure, and grain size distribution. WR stars could be a major contributor to the dust enrichment within galaxies, especially seeding the ISM with small grains (< 0.1 µm). Finally, with this work, we offer a way to trace the physical processes of dust nucleation by only using the geometrical parameters quantifiable using high angular resolution techniques (eg. interferometry, AO imaging). As we hope this paper has illustrated, there is very strong motivation to carefully consider the fascinating WR systems for their deep reach into dust formation physics and for the wider implications into and beyond our own galaxy.

\section*{Acknowledgements}

We acknowledge support from the Australian Research Council (DP 180103408) that funded this work. This project was concluded thanks to the funding from the European Research Council (ERC) under the European Union’s Horizon 2020 research and innovation programme (grant agreement No 742095; {\it SPIDI}: Star-Planets-Inner Disk-Interactions, \url{http://www.spidi-eu.org}). A. L acknowledges funding from the \emph{Agence Nationale de la Recherche} (ANR) in the COSMERGE grant ANR-20-CE31-001 as well as the "Programme National des Hautes Energies" (PNHE) of CNRS/INSU co-funded by CEA and CNES" .
F. M. acknowledges funding from the \emph{Agence Nationale de la Recherche} (ANR) in the MASSIF project ANR-21-CE31-0018-01. This research made use of NASA's Astrophysics Data System; the \textsc{IPython} package \citep{ipython}; \textsc{SciPy} \citep{scipy}; \textsc{NumPy} \citep{numpy}; \textsc{matplotlib} \citep{matplotlib}; and Astropy, a community-developed core Python package for Astronomy \citep{astropy}.




\bibliographystyle{mnras}
\bibliography{Biblio_WR_all} 




\appendix


\bsp	
\label{lastpage}
\end{document}